\documentclass[12pt]{article}
\usepackage[margin=2cm]{geometry}
\usepackage{comment}
\usepackage{amsmath,amssymb,mathtools,graphicx,subfigure,setspace}
\usepackage{cite,float}
\usepackage{slashed} 		 
\usepackage{color}
\makeatother

\newcommand{\be}{\begin{equation}}
\newcommand{\bea}{\begin{eqnarray}}
\newcommand{\eea}{\end{eqnarray}}
\newcommand{\ba}{\begin{array}}
\newcommand{\ea}{\end{array}}
\newcommand{\ee}{\end{equation}}
\newcommand{\bes}{\begin{equation*}}
\newcommand{\beas}{\begin{eqnarray*}}
\newcommand{\eeas}{\end{eqnarray*}}
\newcommand{\bas}{\begin{array*}}
\newcommand{\eas}{\end{array*}}
\newcommand{\ees}{\end{equation*}}
\newcommand{\rd}{\dot{r}}

\newcommand{\R}{\mathbb{R}}

\numberwithin{equation}{section}
\begin{document}
\onehalfspacing
\noindent

\begin{titlepage}
\vspace{10mm}

\vspace*{20mm}
\begin{center}
{\Large {\bf Entanglement Entropy for Singular Surfaces in Hyperscaling violating Theories}\\
}

\vspace*{15mm}
\vspace*{1mm}
{Mohsen Alishahiha${}^a$,   Amin Faraji Astaneh$^{b}$,  Piermarco Fonda${}^c$ and 
 Farzad Omidi${}^{d}$}

 \vspace*{1cm}

{\it ${}^a$ School of Physics, ${}^b$  School of Particles and Accelerators, ${}^d$  School of Astronomy,\\
Institute for Research in Fundamental Sciences (IPM)\\
P.O. Box 19395-5531, Tehran, Iran \\  
${}^c$ SISSA and INFN, via Bonomea 265, 34136, Trieste, Italy\\}
 \vspace*{0.5cm}
{E-mails: {\tt alishah@ipm.ir,  faraji@ipm.ir, piermarco.fonda@sissa.it,\\ farzad@ipm.ir}}%

\vspace*{1cm}
\end{center}

\begin{abstract}
We study the holographic  entanglement entropy for singular surfaces in theories described 
holographically by hyperscaling violating backgrounds. We consider singular surfaces 
consisting of cones or creases in diverse dimensions. The structure of UV divergences of entanglement entropy exhibits new logarithmic terms whose coefficients, being cut-off independent, could be used to define new central charges in the nearly smooth limit. We also show that there is a relation between {these central charges} and the one appearing in the two-point function of the energy-momentum tensor. Finally we examine how this relation is affected by considering {higher-}curvature terms in the gravitational action.

\end{abstract}

\end{titlepage}


\section{Introduction}

It is well known that the central charge of {a two dimensional conformal field theory
 is an important quantity characterizing its behaviour: it is ubiquitous in many expressions such as the
central extension of Virasoro algebra, the two point function of energy-momentum tensor, in the
Weyl anomaly and is the coefficient of the logarithmically divergent term in the entanglement 
entropy \cite{Calabrese:2004eu}. It also appears in the expression of Cardy's formula 
for the  entropy. Actually  the corresponding central charge  may be thought of as a 
measure of the number of degrees of freedom of the theory. 
 Moreover Zamolodchikov's c-theorem in two 
dimensions indicates that in any renormalization group flow connecting two fixed points,
the central charge decreases along the flow, thus indicating that IR fixed points are 
characterized by fewer degrees of freedom.

In higher dimensional conformal field theories the situation is completely
different. First of all the conformal group in higher dimensions does not have a central extension 
and thus it is finite dimensional. 
Moreover the parameter which appears  in the two point function of the energy-momentum tensor 
is not generally related to the one multiplying the Euler density in the  Weyl anomaly in even dimensional
conformal field theories\footnote{ It was conjectured \cite{Cardy:1988cwa} that, in  four  
dimensional  space-times, the coefficient that  multiplies  the Euler density always decreases along RG 
flow and may naturally define an a-theorem in four dimensions. This conjecture has been 
proved in
\cite{Komargodski:2011vj}. }, nor is directly related to the cut-off independent terms of the 
entanglement entropy computed for 
a smooth entangling region. 

Indeed if one  computes  entanglement entropy for a given smooth entangling region in 
a $d+1$ dimensional conformal field theory, one finds \cite{{Srednicki:1993im},{Holzhey:1994we}}
\bea
S_E=\sum_{i=0}^{[\frac{d}{2}]-1} \frac{A_{2i}}{d-2i-1}\;\frac{1}{\varepsilon^{d-2i-1}}+
{\delta_{2[\frac{d}{2}]+1,d}} \;{ A_{2[\frac{d}{2}]}} \log\frac{H}{\varepsilon}+{\rm finite\; terms},
\eea
where $\varepsilon$ is a UV cut off,  $A_i$'s are some constant parameters (in particular $A_0$  is proportional to the area of the enclosed entangling region) and $[x]$ denotes
 the integer part of $x$.  $H$  is a typical scale in the model  which could be 
 the size of entangling region. For an even dimensional field theory (odd $d$ in our notation)  
 the coefficient of the 
logarithmic term, $A_{2[\frac{d}{2}]}$, is a universal constant in the sense that it is 
independent of the UV cut off: in other words it is fixed by the intrinsic  properties of the theory.
Two dimensional CFTs fall in this case since the central charge is indeed a universal quantity.

In general for an even dimensional conformal field theory it can be shown that the coefficient of the universal logarithmic term is given in terms of the Weyl anomaly (see for example \cite{{Ryu:2006ef},{Solodukhin:2008dh},{Casini:2011kv}}). In particular, when the entangling region is a sphere the coefficient is exactly the same as the one multiplying the Euler density. For odd dimensional spacetimes (even $d$) one still has a universal constant term which might provide a generalization of the c-theorem for odd dimensional conformal field theories
 \cite{{Myers:2010xs},{Casini:2006hu}}. 
 
Having  said this, it is  natural to pose the question whether one could find 
further logarithmic divergences in the expression of the entanglement entropy whose coefficients, 
being universal in the sense specified above, could reflect certain intrinsic properties of the  theory 
under consideration. Moreover, if there is such a universal term, it would be interesting to understand 
if any relation between it and other charges of the theory is present. Indeed these questions, for 
some particular cases,  have been addressed in the literature  (see for example \cite{{Drukker:1999zq},{Hirata:2006jx},{Myers:2012vs}}). In particular, it was shown that there is also a logarithmic 
term in three dimensions for sets of entangling regions with non-smooth boundary. In \cite{Fonda:2014cca} it was shown numerically that the same logarithmic term arises for finite-sized entagling regions.
 More precisely, for an 
entangling region with a cusp in three dimensions one has \cite{{Drukker:1999zq},{Hirata:2006jx},
{Myers:2012vs}}
\be
S=S_1\frac{L}{\epsilon}+a(\varphi) \log \epsilon+S_0
\ee
where the cusp is specified by an angle defined such that $\varphi = \pi/2$ corresponds to a smooth
 line. Here  $L$ is the length of the boundary of the entangling region and $S_1$ 
is a constant  which depends on the UV cut off,  while  $a(\varphi)$ and $S_0$  are  universal parameters.

 More recently  based on  early  results  \cite{{Drukker:1999zq}, 
{Hirata:2006jx},{Myers:2012vs}} it was shown that ``the ratio $\frac{a(\varphi)}{C_T}$, where
$C_T$ is the central charge in the stress-energy tensor correlator, is an almost universal quantity''
 \cite{{Bueno:2015rda},{Bueno:2015xda}}(see also \cite{Pang:2015lka}). 
  Indeed it was conjectured in those works that in a generic 
 three dimensional conformal field theory there is a universal 
ratio \cite{Bueno:2015rda} 
\be
\frac{\sigma}{C_T}=\frac{\pi^2}{24},
\ee 
where $\sigma$ is defined through the asymptotic behaviour of $a(\varphi)$, {\it i.e.} $
a(\varphi\rightarrow \pi/2)\approx \sigma\; (\varphi-\pi/2)^2$.

The aim of the present paper is to extend the above consideration {to} higher dimensional 
field theories\footnote{ Holographic  entanglement entropy  for certain singular surfaces  in various 
dimensions has been studied in  \cite{Myers:2012vs}
where it was shown that some specific non-smooth entangling regions exhibit new divergences that 
include logarithmic ones (see  Table1 there).}.
Nonetheless, we will consider cases where the dual field theory 
does not even have conformal symmetry.  
More precisely in this paper we shall explore different logarithmic divergences for the 
entanglement entropy of strongly coupled field theories whose gravitational dual 
are provided by geometries with a 
hyperscaling violating factor\cite{{Kiritsis},{Gouteraux:2011ce}}.  The corresponding 
geometry in $d+2$ dimensions is given by (see  Appendix \ref{App:AppendixA})
\be
\label{metric1}
ds^2=r^{-2\frac{\theta}{d}} \left(-r^{2z}dt^2+r^2\sum_{i=1}^ddx_i^2
+\frac{dr^2}{r^2} \right),
\ee
where the constants $z$ and $\theta$ are  dynamical and  hyperscaling violation exponents, 
respectively. This is the most general geometry which is spatially homogeneous and covariant under 
the following scale transformations
\be
t\rightarrow \lambda^z t, \quad r\rightarrow \lambda^{-1} r, \quad x_i\rightarrow \lambda x_i, \quad  
ds_{d+2} \rightarrow \lambda^{\frac{\theta}{d}} ds_{d+2}.
\ee
Note that with a non-zero $\theta$, the line element is not
invariant under rescalings which in the context of AdS/CFT correspondence  indicates  violations of 
hyperscaling in the dual field theory. 
More precisely, while in $(d+1)$-dimensional theories 
without  hyperscaling violating exponent the entropy  
scales as $T^{d}$ with temperature, in the present case, where the metric has a non-zero 
$\theta$, the entropy scales as $T^{(d-\theta)/z}$ \cite{{Gouteraux:2011ce},{Huijse:2011ef}}. 

Holographic entanglement entropy \cite{{Ryu:2006bv},{RT:2006}} for hyperscaling violating 
geometries has been studied 
in {\it e.g.} \cite{{Dong:2012se},{Ogawa:2011bz},{Alishahiha:2012cm}}. An interesting feature of
 metric (\ref{metric1}) is that for the special value of the hyperscaling violating
exponent $\theta=d-1$, the holographic entanglement entropy shows
a logarithmic violation of the area law\cite{{Ogawa:2011bz},{Huijse:2011ef}}, indicating that the 
background \eqref{metric1} could provide
a gravitational dual for a theory with  an ${\cal O}(N^2)$ Fermi surface, where $N$ is the number of 
degrees of freedom. {Time-dependent} behaviour of holographic entanglement entropy in {Vaidya-hyperscaling} violating {metrics} has also been studied in \cite{{Alishahiha:2014cwa},{Fonda:2014ula}}.

In this paper we will study holographic entanglement entropy in the background \eqref{metric1} 
for an entangling region with the form of $c_n\times \R^{d-n-2}$ where $c_n$ is an $n$ dimensional
cone. We will see that holographic computations indicate the presence of new divergences which could include both log and log${}^2$ terms. Such terms 
could provide a new universal charge for the model. Unlike the Weyl anomaly, this 
charge can be defined in both even and odd dimensional theories. 
 We also note that there is
another quantity, defined in arbitrary dimensions, which is the coefficient entering in the
expression of stress-energy tensor two point function.  Following the ideas in  \cite{Bueno:2015rda}, we investigate whether there is a relation
between these two charges. 
We further show that 
there is a relation between them that remains unchanged even when we add corrections due
to the presence of (certain) higher curvature terms. Therefore it is reasonable to conjecture that
 the relation between these two charges is an intrinsic property of the underling theory. It is worth 
 mentioning that although we will mainly consider a theory with hyperscaling violation, when it comes
 to the comparison of charges we will restrict ourselves to $\theta=0$, though making a
 comment on generic $\theta$.  

The paper is organized as follows. In the next Section we will study entanglement entropy of
an entangling region consisting of an $n$-dimensional cone.  In Section 3
we will compare the results with that of smooth entangling region where we will see 
that the corresponding entanglement entropy for the singular surfaces exhibit new divergent 
terms which include certain logarithmic terms. In Section 4, from the coefficient
of logarithmic divergent terms, we will introduce a new charge for the theory which could be
compared with other central charges in the model. 
The last section is devoted to conclusions.


\section{Entanglement entropy for a higher dimensional cone}

In this section we shall study holographic entanglement entropy on a singular region 
consisting of an $n$ dimensional cone $c_{n}$. To proceed it is convenient to use the following parametrization for the metric in $d+2$ dimensions 
\be 
\label{eq metric hs 1}
ds^2 = \frac{L^2}{r_F^{2\frac{\theta}{d}}}\;\frac{-r^{2(1-z)}dt^2 + dr^2 + d\rho^2 +\rho^2 (d\varphi^2+\sin^2\!\varphi\;
d\Omega^2_{n})+d\vec{x}^2_{d-n-2}}{r^{2(1-\frac{\theta}{d})}}.
\ee
Here $L$ is the radius of curvature of the spacetime and $r_F$ is a dynamical scale. 
Indeed the above  metric could provide a gravitational dual for a strongly coupled field 
theory with hyperscaling violation below the dynamical scale $r_F$\cite{Dong:2012se}.

The entangling region, which we choose to be $c_{n}\times \R^{d-n-2}${, i.e. an $n$-cone extended in $d-n-2$ transverse dimensions}, may be 
parametrized in the following way
\be 
\label{eq hs cusp region}
t={\rm fixed}\;\;\;\;\;\;\;\;\;\; 0 \leq \varphi\leq\Omega\;.
\ee
When $n=0$ the entangling region, which {we call a} crease, will be {delimited} by $-{\Omega}\leq  \varphi\leq{\Omega}$. 

Following \cite{{Ryu:2006bv},{RT:2006}}, in order to compute holographic entanglement entropy 
one needs to minimize the area of a co-dimension two hypersurface in the bulk geometry 
\eqref{eq metric hs 1} whose boundary coincides with the boundary of the entangling region.
Given the symmetry of both the background metric and of the shape of the entangling region,
 we can safely assume that the corresponding co-dimension two hypersurface can be
 described as a function  $r(\rho,\varphi)$ and therefore the 
induced metric on the hypersurface is
\be 
\label{eq metric hs 2}
ds^2_{\textrm{ind}} = \frac{L^2}{r_F^{2\frac{\theta}{d}}}\; \frac{ (1+r'^2) d\rho^2 +(\rho^2 + \rd^2)  d\varphi^2+ 2  r' \rd d\rho 
d\varphi +  \rho^2 \sin^2 \varphi \;d\Omega^2_{n}+d\vec{x}^2_{d-n-2}  }{r^{2(1-\frac{\theta}{d})}}.
\ee
where $r'=\partial_\rho r$ and $\rd=\partial_\varphi r$.  By computing the volume element 
associated to this induced metric we are able to compute the area of the surface, and thus the 
holographic entanglement entropy, as follows
\be\label{AA}
{\cal A}=\epsilon_n \frac{\Omega_{n}V_{d-2-n}L^d}{r_F^\theta}\int d\rho\; d\varphi \;\frac{\rho^{n}\sin^{n}\!\varphi}{r^{d-\theta}}
\;\sqrt{\rho^2(1+r'^2)+\rd^2},
\ee
where $V_{d-n-2}$ is the regularized volume of $\R^{d-n-2}$ space and $\Omega_n$ is the volume 
of the $n-$sphere, ${S}^n$. We introduced $\epsilon_n=1+\delta_{n0}$ to make sure that for $n=0$ 
there is a factor of 2, as for $n=0$ the integral over $\varphi$ still span from $0$ to $\Omega$.

Treating the above area functional as an action for a two dimensional dynamical system,
we just need to solve the equations of motion coming from the variation of  the action to find the profile $r(\rho,\varphi)$.
Note, however, that  since the entangling region is invariant under rescaling  of coordinates, dimensional analysis allows to further constrain the solution to take the form
\be
r(\rho,\varphi) = \rho\; h(\varphi)
\ee
so that $h(\Omega)=0$ and, given {radial} symmetry of the background and of the entangling region, $h'(0)=0$.
To find the area one should then compute the on-shell integral \eqref{AA}. However,  given that the integral is UV-divergent, we have to restrict the integration over the portion of surface $r\geq \varepsilon$, and eventually perform the limit $\varepsilon\to 0$ only after a regularization.
In this regard, the domain $\Sigma_{\varepsilon}$ over which the integration has to be carried out becomes 
\be
\Sigma_{\varepsilon} = 
\left\{ \rho \in [\varepsilon/h_0, H] \;\mathrm{and}\; \varphi \in [0,h^{-1}(\varepsilon/\rho)]
\right\}
\ee
where $h_0\equiv h(0)$ and $H\gg \varepsilon$  is an arbitrarily big cutoff for the length of 
the sides of the singular surfaces. Moreover from the positivity of $r$ it follows 
$h^{-1}(\varepsilon\rho)<\Omega$.

To solve the equation of motion derived  from the action \eqref{AA} it is more 
convenient to consider $\varphi$ as a function of $h$, {\it i.e.} $\varphi=\varphi(h)$. In this notation, setting
$r=\rho h$, the area \eqref{AA} reads
\be \label{eq area functional hs h}
\mathcal{A}=\epsilon_n\frac{\Omega_{n}V_{d-n-2}L^d}{r_F^\theta} \int_{\varepsilon/h_0}^{H} \frac{d\rho}{\rho^{d_\theta-n-1}} \int^{h_0}_{\varepsilon/\rho} d h \frac{ \sin^{n}\!\varphi}{h^{d_\theta}} \sqrt{1+ (1+h^2)\;\varphi'^{2}},
\ee
where $d_\theta\equiv d-\theta$. The equation of motion for $\varphi(h)$ is then 
\be
\label{eq minimal eq phi h}
nh \bigg(\varphi'^2 +\frac{1}{1+h^2}\bigg)\!\cot\varphi\! +\!\varphi' \bigg[ \left(\left(h^2+1\right) d_{\theta }-h^2\right) \varphi'^2+d_{\theta }
-\frac{2 h^2}{(h^2+1)}\bigg]\!-\!h \varphi''\! =\!0,
\ee
For $n=0$ this equation, and the expression for the area \eqref{AA}, simplify significantly, and 
become equivalent to the equation and area functional first studied \textit{e.g.} in \cite{Hirata:2006jx}. 
Indeed in this case the corresponding singular surface is a pure crease $k\times \R^{d-2}$. 

Since \eqref{eq minimal eq phi h} is invariant under $h\to -h$ we have that $\varphi(h)$ is an even function. Therefore, if we want to understand the behaviour of the solution near the boundary,  we can Taylor expand $\varphi(h)$  as  follows
\be 
\label{eq h expansion hs}
\varphi (h) = \sum_{i=0}^{+\infty} {\varphi_{2i}}\; h^{2i},
\ee
so that, by substituting it in \eqref{eq minimal eq phi h}, the solution can be found
 order by order by fixing
the coefficients $\varphi_{2i}$. Indeed for the first three orders one finds
\bea\label{eqphi}
&&\bigg( 2( d_\theta-1)\varphi_2 + n \cot\Omega \bigg)h
\\&&\cr &&+
 \bigg[8\varphi_2^3{d_\theta}+n \cot\Omega  \left(4 \varphi_2^2-{\varphi_2} \cot \Omega -1\right)-
 {\varphi_2} (n+4)+4 {\varphi_4} ({d_\theta}-3)\bigg]h^3\cr &&\cr
&& +
\bigg[-n \left(4 \varphi_2^3-\varphi_2+\varphi_4\right) \cot^2\Omega 
+\varphi_2^3 (8 d_\theta-4 n-8)+48 d_\theta  \varphi_2^2 \varphi_4 +n \left(\varphi_2^2+16 \varphi_2 
\varphi_4+1\right) \cot\Omega \cr &&\cr &&\;\;\;\;\;\;
+n \varphi_2^2  \cot^3\Omega + (n+4)\varphi_2-(n+8)\varphi_4+6(d_\theta-5)
 \varphi_6\bigg]h^5+\cdots
=0. \nonumber
\eea
It is clear from this expression that for $d_\theta=2k+1$ with $k=0,1,\cdots$, the coefficient
$\varphi_{2k+2}$ cannot be fixed by this Taylor series.  In fact when $d_\theta$ is an odd number
one has to modify the expansion by allowing for a non-analytic logarithmic term, as in \cite{Myers:2012vs}.  More 
precisely for generic $d_\theta$ one has
\be\label{phi}
\varphi (h) = \sum_{i=0}^{[\frac{d_\theta}{2}]-1} \varphi_{2i}\; h^{2i}+\varphi_{2[\frac{d_\theta}{2}]}
h^{2[\frac{d_\theta}{2}]}\left(c+\frac{1}{2}\delta_{2[\frac{d_\theta}{2}]+1,d_\theta}
\log h^2\right)+{\cal O}(h^{2[\frac{d_\theta}{2}]+2}),
\ee
where we denote with  $[y]$ the integer part of $y$.
With this Taylor expansion the equation of motion can be solved up to order 
${\cal O}( h^{2[\frac{d_\theta}{2}]})$ which is enough to 
 fix all $\varphi_{2i}$ for $i=1,\cdots, [\frac{d_\theta}{2}]$. Note the constant $c$ in the 
above expansion remains undetermined. The explicit expression 
for the coefficients $\varphi_{2i}$ for the few 
first terms  is presented in the Appendix \ref{App:AppendixB}.

Since the solution is regular at the boundary, we can expand in the same manner the integrand of the area functional
\eqref{eq area functional hs h}  around $h=0$ 
\be
\label{eq area integrand expansion hs}
\frac{\sin^{n}\!\varphi}{h^{d_\theta}} \sqrt{1+(1+h^2)\varphi'^{2}} 
=  \sum_{i=0}^{[\frac{d_\theta}{2}]-1} \frac{a_{2i}}{h^{d_\theta-2i}}+\frac{a_{2[\frac{d_\theta}{2}]}}{h}\delta_{2[\frac{d_\theta}{2}]+1,d_\theta}+{\rm finite\; terms},
\ee
where the coefficients $a_{2i}$ can be expressed in terms of  $\varphi_{2i}$.  
The explicit expression of the coefficients $a_{2i}$  for few first terms 
are presented in Appendix \ref{App:AppendixB}.

To regularize the area functional one may add and subtract the singular terms to make
the integration over $h$ finite. Denoting the regular part of the integrand by $A_{\rm reg}$ the equation \eqref{eq area functional hs h}  reads
\be 
\mathcal{A}=\epsilon_n\frac{\Omega_nV_{d-n-2}L^d}{r_F^\theta}\int_{\varepsilon/h_0}^{H} \frac{d\rho}{\rho^{d_\theta-n-1}} 
\left[
\int^{h_0}_{0} d h\; A_{\rm reg}+ 
\int^{h_0}_{\varepsilon/\rho} d h
\left(\sum_{i=0}^{[\frac{d_\theta}{2}]-1} \frac{a_{2i}}{h^{d_\theta-2i}}+\frac{a_{2[\frac{d_\theta}{2}]}}{h}\delta_{2[\frac{d_\theta}{2}]+1,d_\theta}\right)\right],
\ee
where
\be
A_{\rm reg}=\frac{\sin^{n}\!\varphi}{h^{d_\theta}} \sqrt{1+(1+h^2)\varphi'^{2}} 
-\left(\sum_{i=0}^{[\frac{d_\theta}{2}]-1} \frac{a_{2i}}{h^{d_\theta-2i}}+\frac{a_{2[\frac{d_\theta}{2}]}}{h}\delta_{2[\frac{d_\theta}{2}]+1,d_\theta}\right).
\ee
It is then straightforward to perform the integration over $h$ for the last term. Doing so, one arrives at
\bea \label{FF}
\mathcal{A}\!\!&\!\!=\!\!&\!\!\epsilon_n
\frac{\Omega_nV_{d-n-2}L^d}{r_F^\theta} A_0 \int_{\varepsilon/h_0}^{H}\! 
\frac{d\rho}{\rho^{d_\theta-n-1}}\!  +\epsilon_n
\frac{\Omega_nV_{d-n-2}L^d}{r_F^\theta}\!\int_{\varepsilon/h_0}^{H}\! d\rho A_1(\rho),
\eea
where 
\bea
A_0&=&
\sum_{i=0}^{[\frac{d_\theta}{2}]-1} \frac{-a_{2i}}{(d_\theta-2i-1)h_0^{d_\theta-2i-1}}+a_{2[\frac{d_\theta}{2}]}\delta_{2[\frac{d_\theta}{2}]+1,d_\theta}\log h_0+\int^{h_0}_{0} d h\; A_{\rm reg}\, ,
\cr &&\cr
A_1(\rho)&=&\sum_{i=0}^{[\frac{d_\theta}{2}]-1} \frac{a_{2i}}{(d_\theta-2i-1)}
\frac{\rho^{n-2i}}{\varepsilon^{d_\theta-2i-1}}+a_{2[\frac{d_\theta}{2}]}
\delta_{2[\frac{d_\theta}{2}]+1,d_\theta}\frac{\log \frac{\rho}{\varepsilon}}{\rho^{d_\theta-n-1}}\, .
\eea
In order to evaluate the last integral in the equation \eqref{FF} special care is needed.  Indeed if $n$ 
is an odd number then one may get a logarithmically divergent term from integration over $\rho$
when $i=[\frac{n}{2}]+1$, which may happen only if $[\frac{n}{2}]\leq [\frac{d_\theta}{2}]-2$, 
which can happen only for $d_\theta\geq 4$.  Therefore it is useful to rewrite $A_1(\rho)$ as follows
\bea
A_1(\rho)=
\sum_{i=0}^{[\frac{d_\theta}{2}]-1\;\prime} \frac{a_{2i}}{(d_\theta-2i-1)}
\frac{\rho^{n-2i}}{\varepsilon^{d_\theta-2i-1}}+\delta_{2[\frac{n}{2}]+1,n}
\frac{a_{2[\frac{n}{2}]+2}\;\varepsilon^{3-d_\theta+2[\frac{n}{2}]}}{(d_\theta-2[\frac{n}{2}]-3)\rho}
+
\delta_{2[\frac{d_\theta}{2}]+1,d_\theta}\frac{a_{2[\frac{d_\theta}{2}]}\log \frac{\rho}{\varepsilon}}{\rho^{d_\theta-n-1}},
\eea
where the prime in the summation indicates that when $n$ is an odd number the term at position $i=[\frac{n}{2}]+1$ should be excluded from the sum.  
With this notation and for  $d_\theta-n\neq 2$  one finds
\bea
&&\int_{\varepsilon/h_0}^H d\rho\; A_1(\rho)=
\sum_{i=0}^{[\frac{d_\theta}{2}]-1\;\prime} \frac{a_{2i}}{(n-2i+1)(d_\theta-2i-1)}\left(
\frac{H^{n-2i+1}}{\varepsilon^{d_\theta-2i-1}}-\frac{h_0^{2i-n-1}}{\varepsilon^{d_\theta-n-2}}
\right) \\
&&
\;\;\;\;\;\;\;\;\;\;\;\;\;\;\;\;\;
-\frac{a_{2[\frac{d_\theta}{2}]}
\delta_{2[\frac{d_\theta}{2}]+1,d_\theta}}{(d_\theta-n-2)^2}\left(
\frac{1+ (d_\theta -n-2) \log \left(\frac{H }{\varepsilon }\right)}{
H^{d_\theta -n-2}}-\frac{1- (d_\theta -n-2) \log h_0}{
(\varepsilon/h_0)^{d_\theta -n-2}}\right)\nonumber\\
&&\;\;\;\;\;\;\;\;\;\;\;\;\;\;\;\;\;\;+\delta_{2[\frac{n}{2}]+1,n}\frac{a_{2[\frac{n}{2}]+2}}{(d_\theta-2[\frac{n}{2}]-3)}\;
\frac{\log\frac{Hh_0}{\varepsilon}}{\varepsilon^{d_\theta-2[\frac{n}{2}]-3}}\, .
\nonumber 
 \eea
Moreover from the first term in \eqref{FF} and for $d_\theta-n\neq 2$ one gets
\be
\int_{\varepsilon/h_0}^{H}\! \frac{d\rho}{\rho^{d_\theta-n-1}}
\;=\frac{1}{d_\theta-n-2}\left(\frac{h_0^{d_\theta-n-2}}{\varepsilon^{d_\theta-n-2}}-
\frac{1}{H^{d_\theta-n-2}}\right).
\ee
Altogether the divergent terms of the holographic entanglement entropy for 
$d_\theta\neq n+2$ are obtained
\bea\label{S1}
S\!\!&\!\!=\!\!&\!\!\epsilon_n\frac{\Omega_n V_{d-n-2}L^d}{4Gr_F^\theta}\Bigg[\!
\sum_{i=0}^{[\frac{d_\theta}{2}]-1\;\prime}\!\! \frac{a_{2i}}{(n-2i+1)(d_\theta-2i-1)}\!\left(
\frac{H^{n-2i+1}}{\varepsilon^{d_\theta-2i-1}}-\frac{h_0^{2i-n-1}}{\varepsilon^{d_\theta-n-2}}
\right)\!+\!
\frac{\delta_{2[\frac{n}{2}]+1,n}a_{2[\frac{n}{2}]+2}}{(d_\theta-2[\frac{n}{2}]-3)}\;
\frac{\log\frac{Hh_0}{\varepsilon}}{\varepsilon^{d_\theta-2[\frac{n}{2}]-3}}
\nonumber\\&\!\!+\!\!&
\frac{A_0 }{d_\theta-n-2}
\frac{h_0^{d_\theta-n-2}}{\varepsilon^{d_\theta-n-2}}
\!-\!\frac{a_{2[\frac{d_\theta}{2}]}
\delta_{2[\frac{d_\theta}{2}]+1,d_\theta}}{d_\theta-n-2}
\!\left(\!
\frac{  \log \left(\frac{H }{\varepsilon }\right)}{
H^{d_\theta -n-2}}- 
\frac{1- (d_\theta -n-2) \log h_0}{(d_\theta-n-2)
(\varepsilon/h_0)^{d_\theta -n-2}}\!\right)\!
\!\Bigg]\!+\!{\rm finite\; terms}.
\eea
From this general expression we observe that the holographic entanglement entropy 
for a singular surface shaped as $c_n\times \R^{d-n-2}$ contains many divergent terms 
including, when $d_\theta$ is an odd number\footnote{It is worth noting that although 
the dimension $d$ is an integer number, the hyperscaling violating exponent, $\theta$, does not need to be an integer number. Therefore the effective dimension, $d_\theta$, generally,  may not 
be an integer. For non-integer $d_\theta$ we do not get any universal terms. },  a logarithmically divergent term whose coefficient is universal, in the sense that it is $\varepsilon$ independent. This is the same behaviour for a generic entangling region where in even dimensional CFTs the entanglement entropy contains always a logarithmically divergent term.

On the other hand when $d_\theta=n+2$ the holographic entanglement entropy gets new 
logarithmic divergences. Indeed in this case the last two terms in \eqref{S1} get modified,
leading to
\bea\label{S2}
S\!\!&\!\!=\!\!&\!\!\epsilon_n\frac{\Omega_n V_{d-n-2}L^d}{4G r_F^\theta}\Bigg[\!\!
\sum_{i=0}^{[\frac{d_\theta}{2}]-1\;\prime}\!\!\! \frac{a_{2i}}{(n-2i+1)(d_\theta-2i-1)}\!\left(
\frac{H^{n-2i+1}}{\varepsilon^{d_\theta-2i-1}}-\frac{h_0^{2i-n-1}}{\varepsilon^{d_\theta-n-2}}
\right)\!\!+\!\!
\frac{\delta_{2[\frac{n}{2}]+1,n}a_{2[\frac{n}{2}]+2}}{(d_\theta-2[\frac{n}{2}]-3)}
\frac{\log\frac{Hh_0}{\varepsilon}}{\varepsilon^{d_\theta-2[\frac{n}{2}]-3}}
\nonumber\\&&\;\;\;\;\;\;\;\;\;\;\;\;\;\;\;\;\;\;\;\;\;\;\;\;\;\;\;\;\;\;\;
+A_0\log\frac{Hh_0}{\varepsilon}+
\frac{a_{2[\frac{d_\theta}{2}]}}{2}\;\delta_{2[\frac{d_\theta}{2}]+1,d_\theta}\log^2 \left(\frac{H }{\varepsilon }\right)
\!\Bigg]\!+\!{\rm finite\; terms}.
\eea
It is easy to see that for $\theta=0$ these results reduce to that in \cite{Myers:2012vs}.
In particular  for $\theta=0$ and odd $d$ (even dimension in the notation of  \cite{Myers:2012vs})
where $d=n+2$ one finds a new $\log^2 H/\varepsilon$ divergent term. Comparing 
with the table 1 in \cite{Myers:2012vs} this divergent term appears in background
space-times $\mathbb{R}^4$ and $\mathbb{R}^6$ with cones $c_1$ and $c_3$ respectively. 
For both cases we have $d=n+2$.

It is, however, interesting to note that in the present case the condition to get
squared logarithmic terms is $d_\theta=n+2$  (for $d_\theta \geq 2$) which allows us to have this 
divergent term in any dimension if the hyperscaling violating exponent, $\theta$, is chosen
properly.


\section{New divergences and Universal terms}

In the previous section we have studied  possible divergent terms which could appear in the 
expression for the area of minimal surfaces ending on singular boundary regions.
However, we should be able to distinguish which new logarithmic divergences arise because of the 
singular shape of the entangling region and which arise because of the choice of a non trivial 
hyperscaling violating exponent $\theta$. To this purpose and to isolate the universal terms coming 
from the choice of the shape and not of the background, we study, in this section, the behaviour
 of the divergences in the HEE for a smooth region, and compare with the results of the previous 
 section.

To  find the divergent terms for a smooth surface,  following our notation,
we will parametrize the metric as follows
\be 
ds^2 = \frac{L^2}{r_F^{2\frac{\theta}{d}}}\;\frac{-r^{2(1-z)}dt^2 + dr^2 + d\rho^2 +\rho^2 (d\varphi^2+\sin^2\varphi
d\Omega^2_{n})+d\vec{x}^2_{d-n-2}}{r^{2(1-\frac{\theta}{d})}}.
\ee
We would like  to compute the holographic entanglement entropy for a smooth entangling region  
given by 
\be \label{yy}
t={\rm fixed}\;\;\;\;\;\;\;\;\;\; \rho\leq H,
\ee
with this condition it is clear that the entangling region consists of the direct product between a ball 
and an infinite hyperplane, namely $B^n \times \mathbb{R}^{d-n-2} $.
To compute the entanglement entropy again we should essentially minimize the area 
which in our case is given by
\be\label{oo}
{\cal A}_{\rm smooth}=\frac{\Omega_{n}V_{d-n-2}L^d}{r_F^\theta}\int_0^\pi d\varphi \sin^n\varphi \int dr\; \frac{\rho^{n+1}\sqrt{1+\rho'^2}}{r^{d_\theta}}.
\ee 
Using this expression and following the procedure we have explored in the previous section one 
can find the divergent terms of holographic entanglement entropy for the smooth entangling
surface \eqref{yy} as follows
\be\label{Sm}
S_{\rm smooth}\!=\!\epsilon_n \frac{\sqrt{\pi} \Gamma \left(\frac{n+1}{2}\right)
\Omega_{n}V_{d-n-2}L^d}
{4Gr_F^\theta\Gamma\left(\!\frac{n}{2}+1\right)}\left(\!\!\sum_{i=0}^{[\frac{d_\theta}{2}]-1}\!\!
 \frac{b_{2i}}{d_\theta-2i-1}\frac{1}{\varepsilon^{d_\theta-2i+1}}\!+\!b_{2[\frac{d_\theta}{2}]}
\delta_{2[\frac{d_\theta}{2}]+1,d_\theta}\! \log\frac{H}{\varepsilon}\!\right)\!+\!{\rm finite\; terms}.
\ee
where $b_{2i}$'s are coefficients  appearing in the expansion of the area
\be
 \frac{\rho^{n+1}\sqrt{1+\rho'^2}}{r^{d_\theta}}=\sum_{i=0}^{[\frac{d_\theta}{2}]-1} \frac{b_{2i}}{r^{d_\theta-2i}}+
\delta_{2[\frac{d_\theta}{2}]+1,d_\theta}\frac{b_{2[\frac{d_\theta}{2}]}}{r} \:,
 \ee
which can be found from the equation of motion deduced from \eqref{oo}. 
In particular the coefficient of the universal term for different (odd) $d_\theta$ is found to be
\bea\label{b0s}
d_\theta=1&:& b_0=
H^{n+1},
\cr
d_\theta=3&:& b_2=
-
\frac{(1+n)^2}{8}
 H^{n-1}.
\eea
 Setting $n=d-2$ in the above expressions we find the universal term of the holographic 
 entanglement entropy for a sphere. 
 \\
 We can make another choice of a smooth entangling region, that is an infinite strip (i.e. the product 
 between an interval and an hyperplane). Denoting the width of the strip by $\ell$, 
 the corresponding entanglement entropy for $d_\theta\neq 1$  
 is \cite{{Dong:2012se},{Alishahiha:2012cm}}
  \be\label{SS1}
 S_{\rm smooth}=\frac{L^d V_{d-1}}{4(d_\theta-1)Gr_F^{d-d_\theta}}\left[\frac{2}{\varepsilon^{d_\theta-1}}-\left( \frac{2\sqrt{\pi}\Gamma\left(\frac{d_\theta+1}{2d_\theta}\right)}{\Gamma\left(\frac{1}{2d_\theta}\right)} \right)^{d_\theta}\;\frac{1}{\ell^{d_\theta-1}}\right],
 \ee
 while for $d_\theta=1$ one has
 \be\label{SS2}
 S_{\rm smooth}=\frac{L^dV_{d-1}}{2G r_F^{d-1}}\log\frac{\ell}{\varepsilon}.
 \ee
It is worth noting that when $d_\theta=1$ the leading divergent term is logarithmic, indicating that
the  dual strongly coupled field theory exhibits a Fermi surface \cite{{Huijse:2011ef},
{Ogawa:2011bz}}.

 Comparing these expressions  with equations \eqref{S1} and \eqref{S2} one 
observes that  beside the standard divergences,
there are  new divergent terms due to singular structure of the entangling region.
In particular there are either new log or log${}^2$ terms, whose coefficients are universal in the 
sense that they are independent of the UV cut off. To proceed note that 
for  $d_\theta\neq n+2$ {the} universal term should be read from equation \eqref{S1},
that is
\be\label{U1}
S_{\rm univ}=
-
\delta_{2[\frac{d_\theta}{2}]+1,d_\theta}\;\epsilon_n
\frac{\Omega_n V_{d-n-2}a_{2[\frac{d_\theta}{2}]}L^dH^{n+2-d_\theta}}{4(d_\theta-n-2)\;r_F^\theta
\;G}\log \left(\frac{H }{\varepsilon }\right),
\ee
which is non-zero for odd $d_\theta$.  On the other hand  for $d_\theta= n+2$ the universal
term can be found from \eqref{S2} to be
\be\label{U2}
S_{\rm univ}=\epsilon_n\frac{\Omega_n V_{d-n-2}L^d}{4G\;r_F^\theta}\left[A_0\log\frac{Hh_0}
{\varepsilon}+
\frac{a_{2[\frac{d_\theta}{2}]}}{2}\;\delta_{2[\frac{d_\theta}{2}]+1,d_\theta}\log^2 \left(\frac{H }
{\varepsilon }\right)\right].
\ee
Observe that in this case for any (integer)  $d_\theta$ the first term is always present though  
the log${}^2$
term appears just for odd $d_\theta$. As already noted in \cite{Myers:2012vs}, it is important to note 
that  when $d_\theta$ is
odd the universal term is given by log${}^2$ and the term linear in $\log \varepsilon$ is not universal 
any more.

Using these results one may define the coefficient of the logarithmic term, normalized to the 
volume of the entangling region, as follows
\bea\label{GN}
&&C_{\rm singular}^{\rm EE}=-\epsilon_n\frac{3L^d}{4 (d_\theta-n-2) G }
\;a_{2[\frac{d_\theta}{2}]},\;\;\;\;\;\;\;
{\rm for}\;d_\theta\;{\rm odd,\; and}\;\;d_\theta\neq n+2,\cr
&&C_{\rm singular}^{\rm EE}=-\epsilon_n \frac{3L^d}{ 4G }\;\frac{a_{2[\frac{d_\theta}{2}]}
}{2},\;\;\;\;\;\;\;\;\;\;\;\;\;\;\;\;\;\;\;\;\;\;\;
{\rm for}\;d_\theta\;{\rm odd,\; and}\;\;d_\theta= n+2,\cr
&&C_{\rm singular}^{\rm EE}=-\epsilon_n \frac{3L^d}{ 4G }\;A_0,\;\;\;\;\;\;\;\;
\;\;\;\;\;\;\;\;\;\;\;\;\;\;\;\;\;\;\;
{\rm for}\;d_\theta\;{\rm even,\; and}\;\;d_\theta= n+2,
\eea
where the  explicit form of $A_0$ and $a_{2[\frac{d_\theta}{2}]}$ are given in the previous section
and in the Appendix B. The factor of 3 in the above expressions is due to our normalization, 
which has been fixed by comparing with the entanglement entropy of 2D CFT written as $\frac{c}
{3}\log\ell/\varepsilon$.

Although the general form of the coefficients of the universal terms  
are given in the equation \eqref{GN} it is illustrative to present their explicit forms for 
particular values of $n$ and $d_\theta$. 


\subsection{$d_\theta=1$}

As we have seen the holographic entanglement entropy for a hyperscaling violating metric exhibits 
a log term divergence for $d_\theta=1$ even for a smooth surface. This may be understood 
from the fact that the  underlying dual theory may have a Fermi surface 
\cite{{Huijse:2011ef},{Ogawa:2011bz}}. {For}  $\theta=0$ (that is $d=1$)  {we indeed recover} the logarithmic term of 2D conformal field theories\cite{Calabrese:2004eu}. When 
$\theta\neq 0$ the physics is essentially controlled by the effective dimension
$d_\theta=d-\theta$. Therefore even for higher dimensions $d\geq 2$ with a proper $\theta$ 
such that $d_\theta=1$ the holographic entanglement entropy always exhibit a 
leading logarithmically divergent term.

In this case for an entangling region with a singularity, which clearly is meaningful only 
for $d\geq 2$, using  the explicit expression for  $a_{0}$ one gets
\be
C_{\rm singular}^{\rm EE}=\epsilon_n\frac{3L^d}{4
G}\;\frac{\sin^n\Omega}{n+1},
\ee
while for a smooth surface one has
\be
C_{\rm smooth}^{\rm EE}=\epsilon_n\frac{3L^d}{4G}
\ee
Note that for $n=0$ both charges become the same. Note that
for $n>1$ the coefficient of universal term  $C_{\rm singular}^{\rm EE}$ is smaller than
 the one of the strip by a factor 
of  $\frac{\sin^n\Omega}{2(n+1)}$ and it vanishes in the limit of $\Omega\rightarrow 0$.
 
 \subsection{$d_\theta=2$}
For $d_\theta=2$ being an even number, the holographic entanglement entropy has 
a universal logarithmic term only for $n=0$  which is\cite{PE}
\be
C_{\rm singular}^{\rm EE}=\frac{3L^d}{2G} \;A_0,
\ee
where
\be
A_0=-\frac{1}{h_0}+\int_0^{h_0}dh\left(\frac{\sqrt{1+(1+h^2)\varphi'^2}}{h^2}-\frac{1}{h^2}\right).
\ee
Actually since the expressions we have found  are independent of $\theta$ one may use the results
of $d=2, \theta=0$ to compute the above universal term. Indeed in this case 
one has (see for example \cite{{Hirata:2006jx},{Bueno:2015rda},{Myers:2012vs}})
\bea
C_{\rm singular}^{\rm EE}=\Bigg\{ \begin{array}{rcl}
&\frac{3L^d}{2\pi G} \frac{\Gamma(\frac{3}{4})^4}{\Omega}&\,\,\,\Omega\rightarrow  0,\\ 
&\frac{3L^d}{8\pi G} (\frac{\pi}{2}-\Omega)^2 &\,\,\,\Omega\rightarrow  \frac{\pi}{2}.
\end{array}\,\,
\eea


\subsection{$d_\theta=3$}

In this case when $n\neq 1$ the holographic entanglement entropy has a log term whose
coefficient may be treated as a universal factor given by 
 \be
C_{\rm singular}^{\rm EE}=\frac{3n^2 L^d}{32G}\; \frac{\cos^2\Omega}
{(1-n)\sin^{2-n}\Omega}.
\ee
On the other hand   for $n=1$ the universal term should be read from the log${}^2$ term with the coefficient
\be
C_{\rm singular}^{\rm EE}=\frac{3L^d}{32G}
\frac{\cos^2\Omega}{2\sin\Omega}.
\ee
which in the limit of planar and zero angle behaves as
\bea
C_{\rm singular}^{\rm EE}=\Bigg\{ \begin{array}{rcl}
&\frac{3n^2 L^d}{32 G} \frac{1}{(1-n)\Omega^{2-n}}&\,\,\,\Omega\rightarrow  0,\\ 
&\frac{3n^2 L^d}{32 G} \frac{(\frac{\pi}{2}-\Omega)^2}{1-n} &\,\,\,\Omega\rightarrow  \frac{\pi}{2}.
\end{array}\,\,
\eea
Note that for $n=1$ the factor of $1-n$ in the denominator should be replaced by 2. It is worth 
noting that for $n=0$ the universal charge is zero identically. Therefore for a singular surface 
containing a crease there is not a universal term.


\subsection{$d_\theta=4$}

In this case we get only for  $n=2$ a universal term, which should be read from the equation \eqref{U2}, that is
\be
C^{\rm EE}_{\rm singular}=\frac{3L^d}{4 G } A_0 ,
\ee
where
\be
A_0=\frac{\sin^2\Omega}{3 h_0^3}-\frac{4}{9}\frac{\cos^2\Omega}{h_0}+\int_0^{h_0}dh
\left( \frac{\sin^2\varphi \sqrt{1+(1+h^2)\varphi'^2}}
{h^4}+\frac{\sin^2\Omega}{h^4}-\frac{4}{9}\frac{\cos^2\Omega}{h^2}\right).
\ee
Since we have $n=2$ this result is valid for $d\geq 4$. 

The computation of $A_0$ cannot be performed {analytically}, since we are not able to find a closed expression {for} the profile $h(\varphi)$, however it can still be found numerically.\\
We solved the equation of motion for $\varphi$ and found it as a 
function of $h_0$, thus founding the dependence of $\Omega$ on $h_0$. Then we computed the 
area and by shooting the solution we were able to find $A_0$ as a function of the opening angle 
$\Omega$.
The results are shown in Fig. \ref{f1}.
\begin{figure}[ht]
\begin{center}
\includegraphics[width=.45\textwidth,natwidth=610,natheight=642]{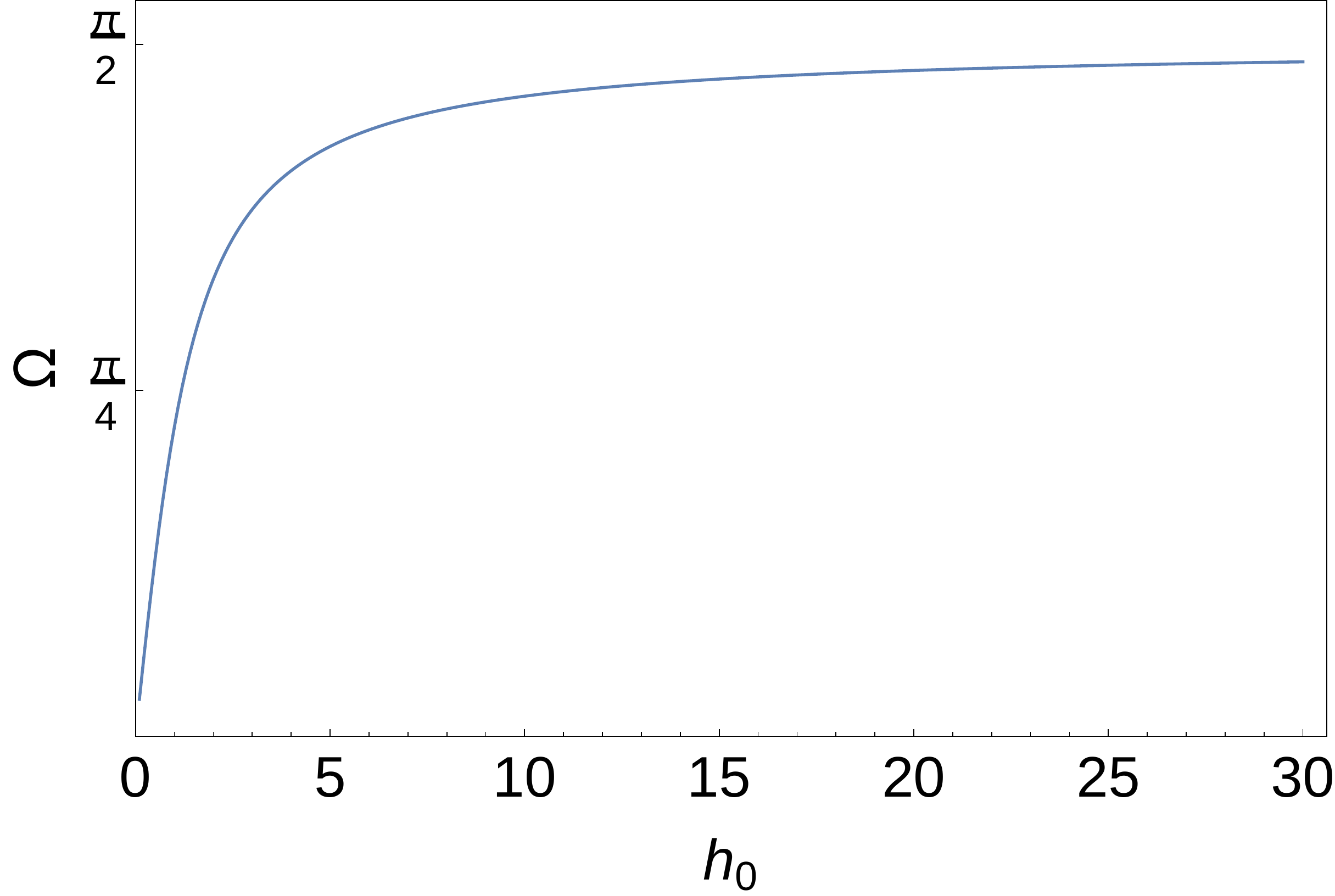}
\includegraphics[width=.442\textwidth,,natwidth=610,natheight=642]{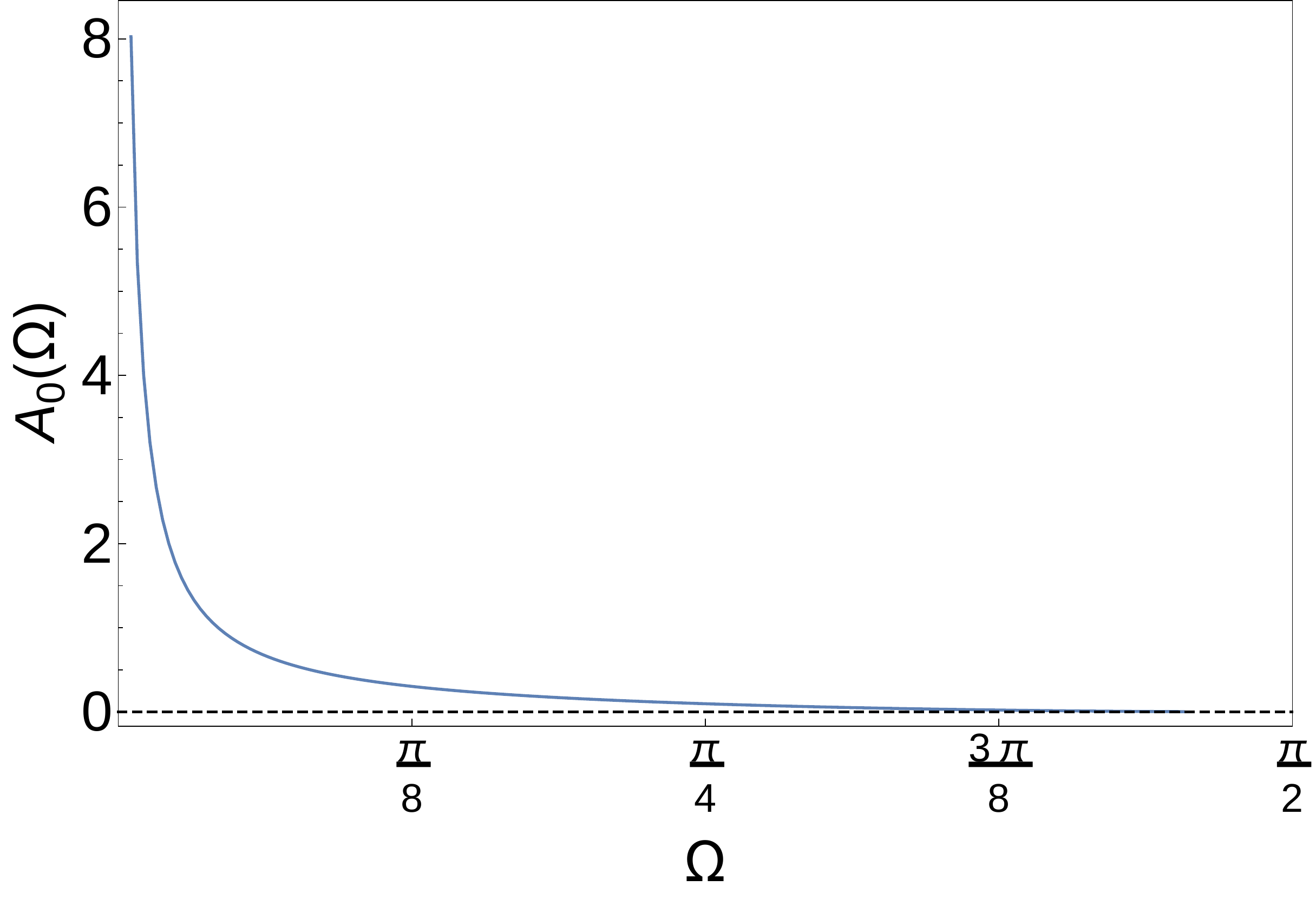}
\caption{$\Omega$ as a function of $h_0$ (left) and  $A_0$ as a function of 
$\Omega$ (right). It shows that  the function $A_0$ diverges at $\Omega=0$ while 
vanishes at $\Omega=\frac{\pi}{2}$.}\label{f1}
\end{center}
\end{figure}
One observes that qualitatively  $A_0$ diverges at $\Omega=0$ while  vanishes at $\pi/2$. 
To make this statement more precise we have numerically 
studied asymptotic behaviours of the function $A_0$ for 
$\Omega\rightarrow 0$ and $\Omega\rightarrow \frac{\pi}{2}$ limits as shown in Fig \ref{f2}.
\begin{figure}[ht]
\begin{center}
\includegraphics[width=.45\textwidth,natwidth=610,natheight=642]{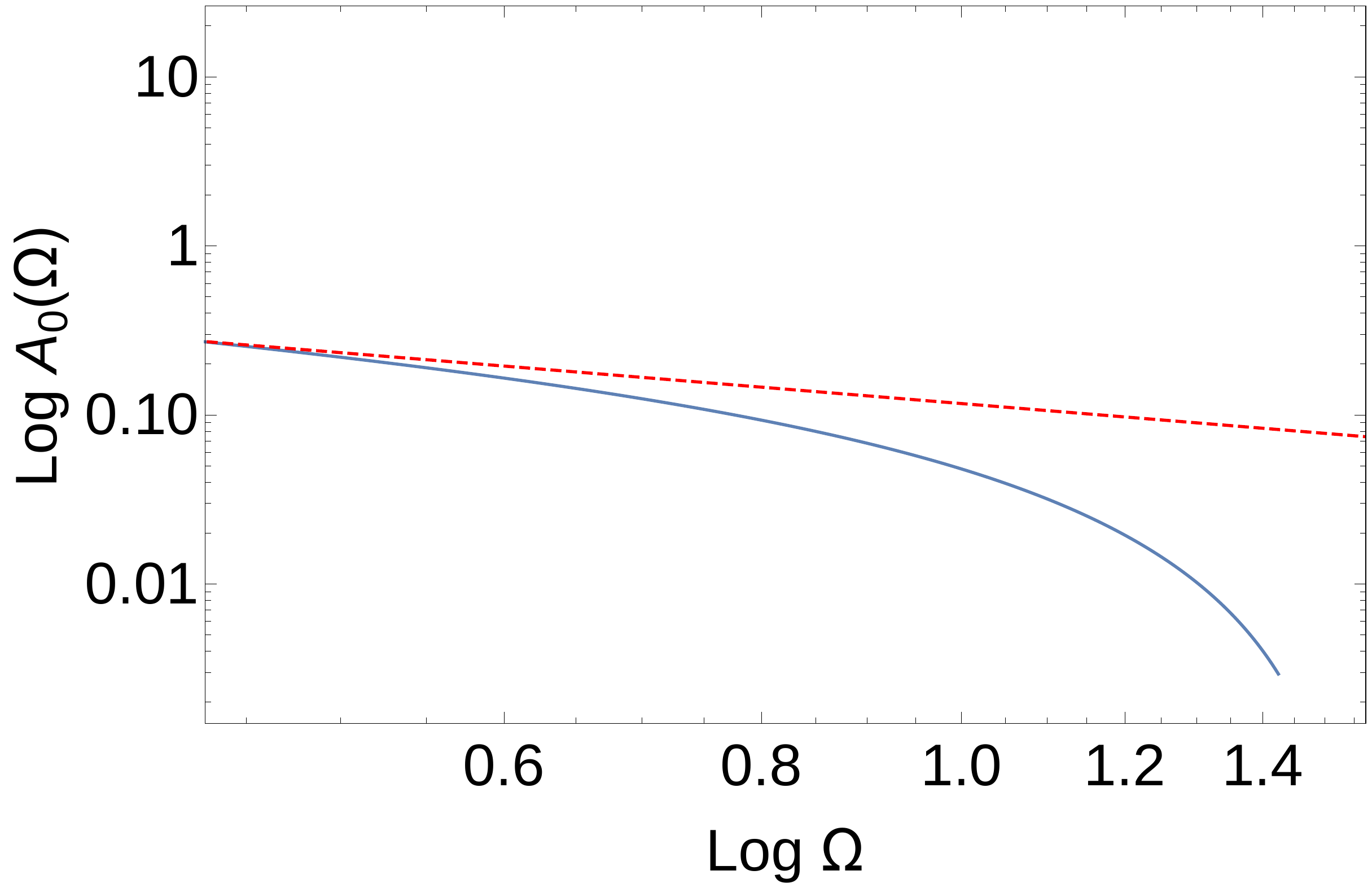}
\includegraphics[width=.45\textwidth,natwidth=610,natheight=642]{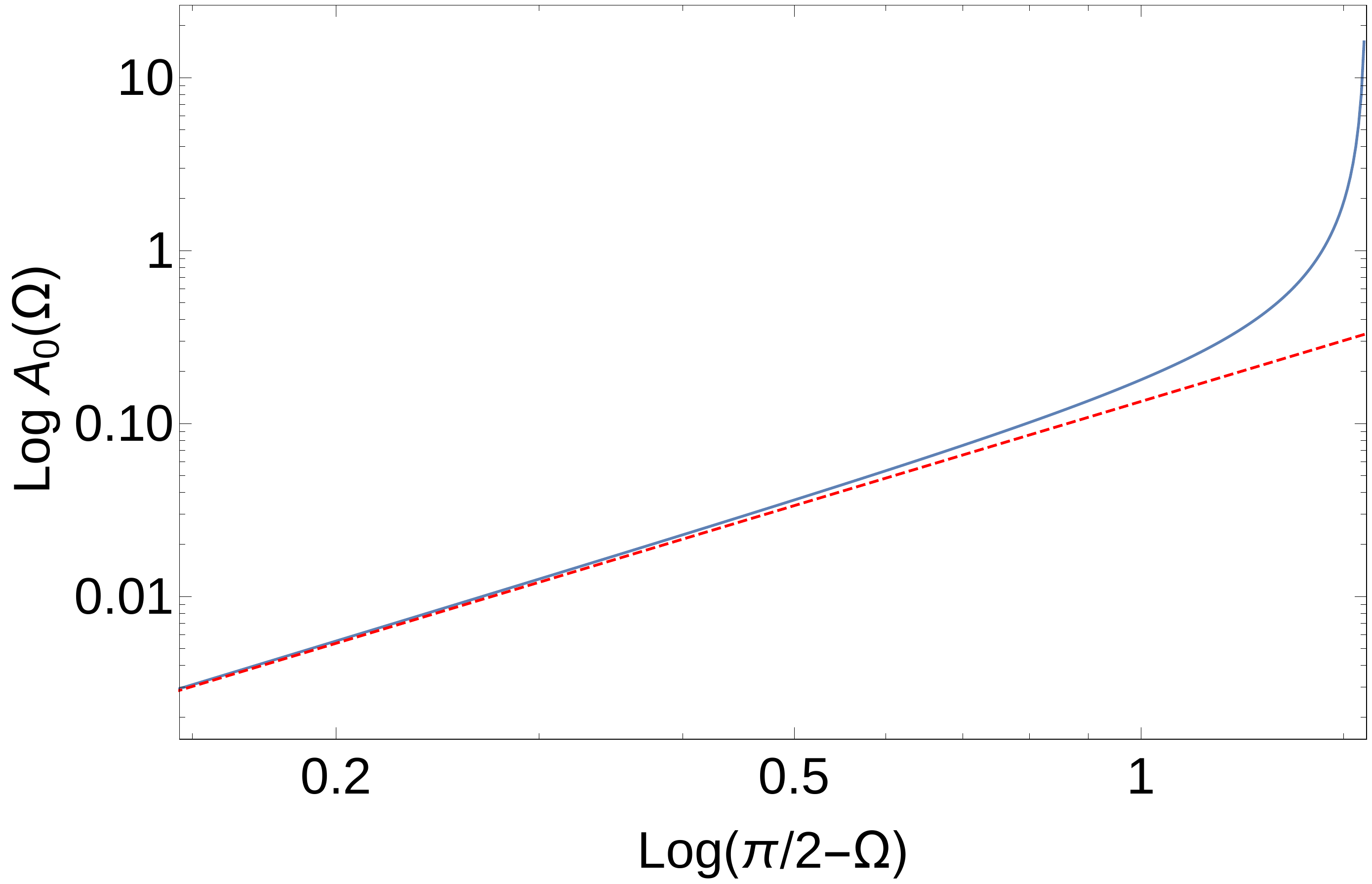}
\caption{Asymptotic behaviours of $A_0$ at $\Omega\rightarrow 0$ (left) and  $
\Omega\rightarrow \frac{\pi}{2}$(right).{ In these plots the  dashed lines 
correspond to test functions to probe the limiting value of $A_0$. The corresponding functions are given by $y= -x-2.15$ (left) and 
$y=2 x-2.01$ (right), in agreement with equation \eqref{ap4}.}  }\label{f2}
\end{center}
\end{figure}
The results may be summarized as follows 
\bea\label{ap4}
C_{\rm singular}^{\rm EE}=\Bigg\{ \begin{array}{rcl}
&\frac{3L^d}{4G}\frac{0.116}{\Omega},&\Omega\rightarrow 0,\\
&\frac{3L^d}{4G} \frac{1.683}{4\pi}\left(\frac{\pi}{2}-\Omega\right)^2, &
\Omega\rightarrow
\frac{\pi}{2}.
\end{array}
\eea   

\subsection{$d_\theta=5$}

In this case and when $n\neq 3$ {we get}
\be
C_{\rm singular}^{\rm EE}=\frac{3n^2 L^d }{4 G}\;\frac{\left(7 n^2-64\right) 
\cos (2 \Omega )+n (7 n-32)+64}{4096(3-n)}\frac{\cos^2\!\Omega}{ \sin^{4-n}\!\Omega}
\ee 
{while for $n=3$}
\be
C_{\rm singular}^{\rm EE}=\frac{3 L^d}{4G}
\frac{9(31-
\cos2\Omega)}{4096}
\frac{\cos^2\Omega}{\sin\Omega}.
\ee
Therefore the corresponding universal term has the following asymptotic behaviours
\bea
C_{\rm singular}^{\rm EE}=\Bigg\{ \begin{array}{rcl}
&\frac{3n^2 L^d }{4G}\;\frac{2n(7n-16)}{4096(3-n)}\frac{1}{ \Omega^{4-n}},&\Omega\rightarrow 0,\\
&\frac{3n^2L^d}{4G} \frac{32(4-n)}{4096(3-n)}\left(\frac{\pi}{2}-\Omega\right)^2, &
\Omega\rightarrow
\frac{\pi}{2},
\end{array}
\eea   
with an obvious replacement for $n=3$.\\ \\

{It is also straightforward to further consider higher $d_\theta$. The lesson we learn 
from these explicit examples is that for a singular surface of the form  $c_n\times \R^{d-n-2}$ and for $d_\theta\geq 2$
 the coefficient of the universal term given in 
the equation \eqref{GN} has the following generic asymptotic behaviour}
\bea
C_{\rm singular}^{\rm EE}\sim\Bigg\{ \begin{array}{rcl}
&\frac{3 L^d }{4G} \frac{1}{ \Omega^{d_\theta-n-1}},&\Omega\rightarrow 0,\\
&\frac{3 L^d}{4G}\left(\frac{\pi}{2}-\Omega\right)^2, &
\Omega\rightarrow
\frac{\pi}{2}.
\end{array}
\eea   
{We see that for a generic opening angle $\Omega$, we can infer the following expression for the coefficient of the universal term}
\be
C_{\rm singular}^{\rm EE}=f_{d_\theta,n}(\Omega)\;\frac{3 L^d}{4G}
\;\frac{\cos^2\!\Omega}{\sin^{d_\theta-n-1}\!\Omega},
\ee
where $f_{d_\theta,n}(\Omega)$ is a function of $\Omega$ which is fixed for given $d_\theta$ and 
$n$ by requiring it to be finite at $\Omega=0$ and $\Omega=\frac{\pi}{2}$.


\section{New charge}

In the previous section we showed that the area of minimal surfaces ending on singular
entangling regions  may present logarithmic divergences for specific choices of the 
extension of the singularity, the dimensionality of the space time and the value of $\theta$. The coefficients of these divergent terms depend on the opening angle of the region, and we were able to compute their value in the nearly smooth limit.
 
Based on these results and using the general  expression given in the equation  \eqref{GN} 
for $d_\theta\geq 2$ one may define a new {\it  charge} as follows
 \be\label{CC}
 C_d^n=\lim_{\Omega\rightarrow \frac{\pi}{2}}\;\frac{C_{\rm singular}^{\rm EE}}{\cos^2\Omega}.
\ee
Note that  this is a well defined limit, {leading to a finite quantity which}  
is proportional to $\frac{L^d}{G }$ up to a numerical factor of order of one. Note also that as soon as we fixed $d_\theta$ the resulting charge is independent of 
$\theta$, and may be defined in any dimension by setting  $n=d_\theta-2$.

As we have already mentioned there is another central charge which
could be defined in any dimension: the coefficient of the $<TT>$ two point function of the stress-energy tensor, which we denote by $C_T$. Following the idea of 
\cite{{Bueno:2015xda},{Bueno:2015rda}}, we can compare these two charges\footnote{Note 
that in even dimensions one may have
another central charge, the coefficient of Euler density  arising in the  computations of the Weyl anomaly. It also
appears as the universal term in the expression of entanglement entropy for a sphere.}. 
Unlike two dimensional CFT where $C_T$ is the same as the one appearing 
in  the central extension of the Virasoro algebra, in higher dimensions  it should be read from the
explicit expression of the two point function. 
Indeed, in the present context, the corresponding two point function may 
be found from the quadratic on-shell action of the perturbation of the metric above a vacuum 
solution  using holographic renormalization techniques\cite{Skenderis:2002wp}.

We note, however, that since we do not have a well defined  asymptotic behaviour of 
metric \eqref{sol} in the sense of Fefferman-Graham expansion, in general it is not an easy task to 
compute the stress-energy tensor's two point function for spacetimes with generic $\theta$ and $z$.
Nevertheless  setting  $z=1$, where one recovers the  Lorentz invariance, we can still use the 
holographic renormalization procedure to find (see Appendix \ref{App:AppendixA})
\be
C_T=\frac{L^d}{8\pi G}\;\;\frac{d+2}
{d}\;\frac{\Gamma(d_\theta+2)}{\pi^{\frac{d+1}{2}}\Gamma\left( 
\frac{1+2d_\theta-d}{2}\right)}.
\ee  
Note that for $z=1$, from the null energy condition 
one gets $\theta(d-\theta)\leq 0$ which has only a partial overlap with the parameter space of 
the model we are considering at $\theta=0$. Therefore using the above expression we 
really should only compare it with the new central charge of the model for $\theta=0$. \\
Since however the new charge defined in \eqref{CC} for given 
$d_\theta$ is independent of $\theta$, the comparison still makes sense. In particular 
for $d_\theta=2,3$ and $d_\theta=4$, respectively,  one finds\footnote{Due to our
 normalization of $C_d$ for $d_\theta=d=2$  there is factor $\frac{1}{3}$ mismatch with the result of \cite{Bueno:2015xda}.}:
\be
\label{eq ratio univ 1}
\frac{C_d^0}{C_T}=\frac{d\;\pi^{\frac{d+1}{2}}\Gamma\left( 
\frac{5-d}{2}\right)}{2(d+2)},\;\;\;\;\;\;\;\;\;\;\;\;\;\;
\frac{C_d^1}{C_T}=-\frac{d\;\pi^{\frac{d+3}{2}}\Gamma\left( 
\frac{7-d}{2}\right)}{64(d+2)},\;\;\;\;\;\;\;\;\;\;
\frac{C_d^2}{C_T}=1.683 \frac{d\;\pi^{\frac{d+1}{2}}\Gamma\left( 
\frac{9-d}{2}\right)}{80(d+2)}
  \ee
For $z\neq 1$, $C_T$ depends explicitly on $z$ and 
thus the above ratio will be $z$ {dependent}, even though $C^n_d$ will not.

Since both central charges considered above are proportional to $\frac{L^d}{G}$, 
it is evident that their ratio is a purely numerical constant. In \cite{Bueno:2015xda} it was conjectured that for three dimensional CFTs this ratio could be completely universal, regardless of the strength of the coupling so to hold in both known statistical models and in QFTs with gravity duals. It is thus interesting to understand whether this ratio, which could characterize whatsoever CFT of fixed dimensionality, is still universal even in the higher dimensional cases we are considering.\\
The easiest step we can make in this direction is to look at gravity theories with higher curvature terms in the action, and see {whether} the corrections alter the ratio \eqref{eq ratio univ 1} .\\

To proceed let us consider an action containing the most general curvature squared
corrections  as follows
\be\label{ACT2}
I=-\frac{1}{16\pi G}\int d^{d+2}\sqrt{-g}\bigg(R+V(\phi)+\lambda_1 R^2+\lambda_2 R_{\mu\nu}
R^{\mu\nu}+\lambda_3 R_{\mu\nu\rho\sigma}R^{\mu\nu\rho\sigma}\bigg)+I_{\rm matter}
\ee
where $I_{\rm matter}$ is a proper matter action to make sure that the model admits hyperscaling 
violating geometry. 
It is then straightforward, although lengthy, to compute holographic entanglement entropy 
for this model\footnote{Holographic entanglement entropy for a strip entangling region in 
theories with hyperscaling violation in the presence of  higher curvature terms has also been studied in \cite{Bueno:2014oua}.}.
 Indeed following \cite{Fursaev:2013fta}, the holographic entanglement entropy   
may be obtained by minimizing the following  entropy functional 
\begin{eqnarray}\label{EE}
S_A\!\!=\!\!\frac{ 1}{4G}\! \int \!d^d\zeta \;\sqrt{\gamma}\;\bigg[1+2\lambda_1 R+\lambda_2 \left({ R}_{\mu\nu}n^\mu_i n^\nu_i-\frac{1}{2}\mathcal{K}^i\mathcal{K}_i\right)\!+\!2\lambda_3 \bigg( R_{\mu\nu\rho\sigma}n^\mu_i n^\nu_j n^\rho_i n^\sigma_j-\mathcal{K}^i_{\mu\nu}\mathcal{K}_i^{\mu\nu}\bigg)\!\bigg],
\end{eqnarray}
where with $i=1,2$ we denote the two transverse directions to a co-dimension two 
hypersurface in the bulk,
$n_i^\mu$  are two mutually orthogonal unit vectors to the hypersurface 
and ${\cal K}^{(i)}$ are the traces of two extrinsic curvature tensors defined by
\be \label{ECUR}
{\cal K}_{\mu\nu}^{(i)}=\pi^\sigma_{\ \mu} \pi^\rho_{\ \nu}\nabla_\rho(n_i)_\sigma,\;\;\;\;\;\;\;
{\rm with}\;\;\;\;  \pi^\sigma_{\ \mu}=\delta^\sigma_{\ \mu}+\xi\sum_{i=1,2}(n_i)^\sigma(n_i)_\mu\ ,
\ee
where $\xi=-1$ for space-like and $\xi=1$ for time-like vectors. Moreover  $\gamma$ is the 
induced metric on the hypersurface whose coordinates are denoted by $\zeta$. 

Although so far we have been considering a theory with hyperscaling violation, as we have already 
mentioned the holographic renormalization for generic hyperscaling exponent has not 
been fully worked out and thus we have restricted ourselves to 
to consider backgrounds with $z=1$. In this case the most interesting case allowed by the null energy condition is $\theta=0$. 
Therefore in what follows we just examine the relation between {the} two charges for $\theta=0$ in an 
arbitrary dimension.
 
To compute higher curvature corrections to the entanglement entropy we note that 
in  our case the normal vectors are given by (note that we set $\theta=0$)
\be
n_1=\frac{L}{r}\bigg(1,0,0,0\cdots\bigg)\,,\;\;\;\;\;\;\;
n_2=\frac{L}{r}\frac{1}{\sqrt{1+h(\varphi)^2+h'(\varphi)^2}}
\bigg(0,1,-h(\varphi),-\rho h'(\varphi),0,\cdots\bigg)\,.
\ee
It is then straightforward to extremize the functional (4.5) and evaluate it.
In fact one only needs to expand the above entropy functional
around $h=0$ to find its divergences and read the universal 
coefficient of the logarithmic (or log${}^2$) term to find the corrections to the central charge
$C_d^n$. Doing so one arrives at
\be
\tilde{C}_d^{\,n}=\Upsilon\; C_d^{\, n}\,,
\ee  
where $\tilde{C}$  is the corrected central charge and 
\be \label{UP}
\Upsilon=1+\frac{4(d-2)}{L^2}\lambda_3-\frac{2(d+1)}{L^2}(\lambda_2+(d+2)\lambda_1)\, .
\ee

Now one needs to compute the corresponding corrections to the central charge $C_T$. To do
so one first needs to linearize the equations of motion deduced from 
the action \eqref{ACT2} (see for example \cite{Gullu:2009vy})
\bea
&&R_{\mu\nu}-\frac{1}{2}g_{\mu\nu}(R+V(\phi))+2\lambda_1\left(R_{\mu\nu}-\frac{1}{4}
g_{\mu\nu}R\right)R+2\lambda_2\left(R_{\mu\sigma\nu\rho}-\frac{1}{4}
g_{\mu\nu}R_{\sigma\rho}\right)R^{\sigma\rho}
\cr &&
+(2\lambda_1+\lambda_2+2\lambda_3)\bigg(g_{\mu\nu}\Box-\nabla_\mu\nabla_\nu\bigg)R
+(\lambda_2+4\lambda_3)\;\Box\left(R_{\mu\nu}-\frac{1}{2}
g_{\mu\nu}R\right)
\cr &&
+2\lambda_3\left(2R_{\mu\sigma\nu\rho}R^{\sigma\rho}+
R_{\mu\sigma\rho\tau}R_{\nu}^{\sigma\rho\tau}-2R_{\mu\sigma}R^{\sigma}_\nu+
\frac{1}{4}g_{\mu\nu}(R_{\alpha\beta\rho\sigma}^2+4R_{\alpha\beta}^2)\right)=0
\eea
Using the notation of Appendix A one can linearize the above equations around the vacuum 
solution given by \eqref{sol} with  $\theta=0$. The result is 
\be
\Upsilon\; {\cal G}_{\mu\nu}^{(1)}+(2\lambda_1+\lambda_2+2\lambda_3)\!\left(\!
\bar{g}_{\mu\nu}\bar{\Box}-\bar{\nabla}_\mu\bar{\nabla}_\nu-\frac{d+1}{L^2}\bar{g}_{\mu\nu}\!\right)\!
R^{(1)}+(\lambda_2+4\lambda_3)\!\left(\!(\bar{\Box}+\frac{2}{L^2}){\cal G}_{\mu\nu}^{(1)}+\frac{d}{L^2}
\bar{g}_{\mu\nu}R^{(1)}\!\right)\!\!=\!0,
\ee
where $\Upsilon$ is exactly the one given in equation \eqref{UP}, and
\be
{\cal G}_{\mu\nu}^{(1)}=R_{\mu\nu}^{(1)}-\frac{1}{2}\bar{g}_{\mu\nu}R^{(1)}+\frac{d+1}{L^2}h_{\mu\nu}.
\ee
In the transverse-traceless gauge  the above equation reads
\be
\left[\Upsilon+(\lambda_2+4\lambda_3)\!
\left(\bar{\Box}+\frac{2}{L^2}\right)\right]\left(\bar{\Box}+\frac{2}{L^2}\right)h_{\mu\nu}=0
\ee
which has to be solved in order to find the linearized solution. Since we are interested in 
the correlation function of the energy momentum tensor, we should still look for a solution 
of $(\bar{\Box}+\frac{2}{L^2})h_{\mu\nu}=0$.
This equation is exactly the same one gets from purely Einstein gravity, and thus the linearized equation of motion reduces essentially to solving standard linearized Einstein equations. {On the 
other hand, to evaluate the} two point function one needs to find the quadratic action which has 
an effective Newton constant $\Upsilon/G$. Indeed going through the computations of the two point function one finally finds that 
\be
\tilde{C}_T=\Upsilon\; C_T,
\ee
and thus we may conclude that
\be
\frac{\tilde{C}_d^{\,n}}{\tilde{C}_T}= \frac{{C}_d^{\,n}}{{C}_T}.
\ee
for arbitrary dimensions but with $\theta=0$.\\
Although we have examined the relation between {the} two central charges $C_T$ and $C_d^{\,n}$ {just for squared} curvature {modifications of Einstein gravity}, based on our observations and {the three-dimensional results of  \cite{Bueno:2015xda}}, it is tempting to conjecture that the the central charge 
$C_d^{\,n}$ 
is directly related to $C_T$ for a generic CFT.

\section{Conclusions}

In this paper we have studied the holographic entanglement entropy of an entangling region  
$c_n\times \mathbb{R}^{d-n-2}$,
i.e. an $n$-dimensional cone extended in $d-n-2$ transverse directions, for a
$d+1$ dimensional theory in a hyperscaling violating background. We have observed that due to the
presence of a corner in the entangling region the divergence structure of the entropy gets new terms.

In particular for certain values of $\theta, d$ and $n$ the divergent terms include log or log-squared
terms whose coefficients are universal, in the sense that they are independent of the UV cut off.

Given that we have been able to extract new regularization independent quantities, 
it is tempting  to conjecture that  some information can be obtained about the underlying dual field theory. This might be compared with the  case of two dimensional conformal field theories where the central charge appears in the coefficient of the (leading) logarithmic divergence of the entanglement entropy for an interval.  

Motivated by this similarity we proceed by analogy and, 
denoting the coefficient of the logarithmic term appearing in the expression for the entanglement entropy by $C_{\rm singular}^{EE}$  (see equation \eqref{GN}), we find that for $d_\theta\geq 2$ we can define a new "central charge" as follows
\be
C_d^{\,n}=\lim_{\Omega\rightarrow \frac{\pi}{2}}\;\frac{C_{\rm singular}^{\rm EE}}{\cos^2\Omega},
\ee
which is proportional to $L^d/G$.
As soon as the effective dimension $d_\theta$ is fixed, the proportionality constant only 
depends on $d$ and $n$, while it is independent of $\theta$. Therefore it remains 
unchanged even if we set $\theta=0${, reducing} the dual theory to 
a $d+1$ dimensional conformal field theory. It is natural to expect that this central charge may 
provide a measure for the number of degrees of freedom of the theory. Note that, unlike the
one obtained from Weyl anomaly, this central charge can be defined for both even and odd dimensions when $
d_\theta=n+2$.

Another central charge which could be defined in any dimension is the one entering 
in the expression for the stress-energy tensor'  two point function.  We checked whether the ratio between 
these charges is a pure number and we also have computed
corrections to both $C_d^n$ and $C_T$ for theories with quadratic correction in the curvature.
 We have shown that the relation between these 
two charges remains unchanged.

Based on this observation and the results for three dimensional CFTs
\cite{{Bueno:2015rda},{Bueno:2015xda}}, one may conjecture  that the relation between these
two central charges ( $C_T$ and $C_d^n$)  is a somehow intrinsic property of the field theory. In fact this relation is reminiscent of the relation between Weyl anomaly 
of a conformal field theory in even dimension and the logarithmic term in the
entanglement entropy of the corresponding theory. If there is, indeed, such a relation one would 
expect to have a general proof for it independently of an explicit example\footnote{ M. A. would like to thank S. Trivedi for 
a discussion on this point.}\cite{TBA}. 



\section*{Acknowledgements}

We would like to thank  A. Mollabashi, M. R. Mohammadi Mozaffar,  A. Naseh, M. R. Tanhayi 
and E. Tonni for useful discussions.
We also acknowledge the use of M. Headrick's excellent Mathematica package "diffgeo". We would 
like to thank him for his generosity.  This work was first presented in {Strings} 2015 and M. A. 
would like to thank the organizers of Strings 2015 for very warm hospitality. M. A. would also like 
to thank S. Trivedi for a discussion.
P.F. would like to thank IPM for great hospitality during part of this project.
This work is supported by Iran National Science Foundation (INSF).

 
\section*{Appendix}
\appendix
\section{Backgrounds with hyperscaling violating factor } \label{App:AppendixA}

In this section we will review certain features of  gravitational backgrounds with 
hyperscaling violating factor\cite{{Kiritsis},{Gouteraux:2011ce},{Dong:2012se}}. 
In what follows we will follow the notation of  \cite{Alishahiha:2012qu} and consider a 
minimal {dilaton}-Einstein-Maxwell action, that is 
\be
\label{action}
S=-\frac{1}{16\pi G}\int d^{d+2}x\sqrt{-g}\left[R-\frac{1}{2}(\partial\phi)^2+V(\phi)-\frac{1}{4}
 e^{\lambda \phi}F_{\mu\nu}F^{\mu\nu}\right], 
\ee
where, motivated by the typical exponential potentials of string {theories}, we will consider the 
following potential
\be
V=V_0e^{\gamma\phi}.
\ee
The equations of motion of the above  action read
\bea
\label{eqmotion1}
&&R_{\mu\nu}+\frac{V(\phi)}{d}g_{\mu\nu}=\frac{1}{2}\partial_\mu\phi\partial_\nu\phi
+\frac{1}{2}e^{\lambda \phi}\left(F_{\mu}^\rho F_{\rho\nu}-\frac{g_{\mu\nu}}{2d}F^2\right),\cr &&\cr
&& \nabla^2\phi=-\frac{dV(\phi)}{d\phi}+\frac{1}{4}\lambda e^{\lambda\phi} 
F^2,\;\;\;\;\;\;
\partial_\mu\left(\sqrt{-g}e^{\lambda\phi}F^{\mu\nu}\right)=0.
\eea
It is straightforward to find a solution to these equation, namely the black brane
\bea\label{sol}
&&ds^2=\frac{L^2}{r^2}\;\left(\frac{r}{r_F}\right)^{2\frac{\theta}{d}}\left(-\frac{f (r)dt^2}{r^{2(z-1)}}+
\frac{dr^2}{f(r)}+d\vec{x}_d^2\right),\;\;\;\;\;\;\;\;f(r)=1-m\; r^{d_\theta+z},\cr&&\cr
&&F_{tr}=\sqrt{2(z-1)(d_\theta+z)}
r^{d_\theta+z-1}, \;\;\;\;\;\;\;\;\;\;\;\;\;\;\;\;\;\;\;\;\;\;\;\;\phi=\sqrt{2d_\theta (z-1-{\theta}/{d})} \log r.
\eea
which solve \eqref{eqmotion1} if we choose the parameters in the action to be
\be
V=\frac{(d_\theta+z)(d_\theta+z-1)}{L^2} \left(\frac{r_F}{r}\right)^{\frac{2 \theta }{d}}\!,\;\;\;\;\;
\lambda=-2\frac{\theta+dd_\theta}{\sqrt{2d d_\theta(dz-d-\theta)}},\;\;\;\;\;
\gamma=\frac{2\theta}{d\sqrt{2d_\theta (z-1-{\theta}/{d})}} .
\ee
Here $L$ is the radius of curvature of the spacetime and $r_F$ is a scale which can be 
interpreted as the gravitational dual of the Fermi radius of the theory living on the boundary.
A charged black brane solution would need more gauge fields to support its 
charge, although in what follows we restrict ourselves to the neutral background. 

This geometry  is a black brane background whose Hawking temperature is 
\be
T=\frac{d_\theta+z}{4\pi\; r_H^z},
\ee
where $r_H$ is the  horizon radius  defined by  $f(r_H)=0$. In terms of the Hawking temperature the thermal entropy can be computed to be
\be
S_{\rm th}=\left(\frac{4\pi}{d_\theta+z}\right)^{\frac{d_\theta}{z}}\;
\frac{L^dV_{d}}{4G\;r_F^{d-d_\theta}}\; T^{\frac{d_\theta}{z}}.
\ee

It is also interesting to evaluate the quadratic  action for a small perturbation above the vacuum solution
\eqref{sol}. This may be used to compute two point function of the energy momentum tensor.
To proceed  we will  consider {a perturbation over} the vacuum in which {we let vary only the metric}
\be
g_{\mu\nu}=\bar{g}_{\mu\nu}+h_{\mu\nu},\;\;\;\;\;\;\;\phi=\bar{\phi},\;\;\;\;\;\;\;\;\;A_\mu=\bar{A}_\mu.
\ee
where the ``bar'' quantities represent the vacuum solution \eqref{sol}. It is then straightforward 
to linearize the equations of motion{, leading to}
\be
R_{\mu\nu}^{(1)}+\frac{V(\bar{\phi})}{d}h_{\mu\nu}=0,\;\;\;\;\;\;\;\;\frac{1}{\sqrt{\bar{g}}}
\partial_\mu\left(\sqrt{\bar{g}} h^{\mu\nu}\partial_\nu\bar{\phi}\right)=\frac{1}{2}\bar{g}^{\mu\nu}\partial_\mu h
\partial_\nu \bar{\phi},\;\;\;\;\;\;\;\;\;\bar{F}^{\mu\nu}\partial_\mu h=0.
\ee 
Here the linearized Ricci tensor is given by
\bea
R_{\mu\nu}^{(1)}&=&\frac{1}{2}\left(-\bar{\nabla}^2 h_{\mu\nu}-\bar{\nabla}_\mu \bar{\nabla}_\nu
h+\bar{\nabla}_\sigma \bar{\nabla}_\nu h^\sigma_\mu+\bar{\nabla}_\sigma \bar{\nabla}_\mu
h^\sigma_\nu\right)\\
&=&\frac{1}{2}\left(-\bar{\nabla}^2 h_{\mu\nu}-\bar{\nabla}_\mu \bar{\nabla}_\nu
h+\bar{\nabla}_\nu \bar{\nabla}_\sigma h^\sigma_\mu+\bar{\nabla}_\mu\bar{\nabla}_\sigma
h^\sigma_\nu+\bar{R}_{\sigma\nu}h^\sigma_\mu+\bar{R}_{\sigma\mu}h^\sigma_\nu-
2\bar{R}_{\lambda\mu\sigma\nu}h^{\lambda\sigma}\right).\nonumber
\eea
Moreover for the Ricci scalar one gets
\be
R^{(1)}=\bar{g}^{\mu\nu}R^{(1)}_{\mu\nu}-\bar{R}_{\mu\nu} h^{\mu\nu}=-\bar{\nabla}^2 h
+\bar{\nabla}_\mu\bar{\nabla}_\nu h^{\mu\nu}-\bar{R}_{\mu\nu} h^{\mu\nu}.
\ee
{In order to solve the equations of motion one needs to properly fix the gauge freedom. In our case it turns out to be useful to choose a covariant gauge $\nabla^\mu h_{\mu\nu}=\frac{1}{2}\nabla_\nu h$, which however still does not fix all redundant degrees of freedom.
Indeed, we fix the remaining ones by setting $h_{ri}=h=0$
and thus $\nabla^\mu h_{\mu\nu}=0$ so that we reduce to a
transverse and traceless gauge.} {It is easy to see, with this constraint and gauge 
choice, that} the equation of
motion of the scalar field at first order will be  identically satisfied and one only needs to solve 
the Einstein equations, which, generally, reduce to {further} equation of motion for a scalar field.
Indeed taking into account that
\be
\bar{R}_{\mu\sigma}h^\sigma_\nu=-\frac{1}{d}\left(V(\bar{\phi})+\frac{1}{4}e^{\lambda\bar{\phi}}
\bar{F}^2 \right)h_{\mu\nu}+\frac{1}{2}h^\sigma_\nu\left(\partial_\mu\bar{\phi}\partial_\sigma
\bar{\phi}+e^{\lambda\bar{\phi}} \bar{F}^\rho_\mu \bar{F}_{\rho\sigma}\right)
\ee
and using the transverse-traceless gauge {we arrive at}
\be
\label{eq grav mot}
\bar{\nabla}^2h_{\mu\nu}+2\bar{R}_{\alpha\mu\beta\nu}h^{\alpha\beta}
+\frac{1}{2d}e^{\lambda \bar{\phi}}
\bar{F}^2 h_{\mu\nu}-\frac{1}{2}e^{\lambda \bar{\phi}}\bar{F}_{\rho\sigma}\bar{F}^\rho_{(\mu} 
h_{\nu)}^\sigma=0.
\ee
Using the parameters of the vacuum {solution, one could} in principle solve the above
differential equations with given boundary condition. Then  by making use of AdS/CFT 
correspondence from the quadratic action one can compute the two point function of 
the energy 
momentum tensor for a strongly coupled field theory whose gravitational dual 
is provided by a geometry with hyperscaling violating factor using 
holographic renormalization. 
\\
In general \eqref{eq grav mot} cannot be solved analytically, and since for $z \neq1$ we do not have a good control on the asymptotic behaviour of the metric (in ananlogy with the Fefferman-Graham expansion), it is hard to use holographic renormalization techniques (see however
\cite{Taylor:2008tg} for a related issue).

On the other hand, setting $z=1$, and thus recovering Lorentz symmetry in the bulk metric, we can rely on the holographic renormalization to compute the stress-energy {tensor} two point's function, namely because the action reduces to a dilaton-Einstein model with a simpler equation of motion
\be\label{EG}
\bar{\nabla}^2h_{\mu\nu}+2\bar{R}_{\alpha\mu\beta\nu}h^{\alpha\beta}=0.
\ee
It is however important to note that the null energy condition for $z=1$ implies that $\theta (d-\theta)\leq 0$, that is either $\theta\leq 0$ or $\theta\geq d$. In all our computations we implicitly assumed $d_\theta\geq 1$, playing $d_\theta$ the role of the effective dimension, although a solution with
 $\theta>d$ may not be consistent\cite{Dong:2012se}.

Moreover, for $\theta=0$ it is clear that all equations reduce to that of Einstein gravity. In particular 
one gets \cite{Liu:1998bu} 
\be
h_k^l(r,x)=\frac{\Gamma\left(d+1\right)}{\pi^{\frac{d+1}{2}}
\Gamma\left(\frac{1+d}{2}\right)}\int dy^{d+1}
\left( \frac{r}{r^2+(x-y)^2}   \right)^{d+1}J^i_k (x-y) J^l_j(x-y)P_{ijab} \;h_{ab}(y),
\ee 
where $h_{ab}$ is the boundary value of metric and ({see \cite{Liu:1998bu}})
\be
J_i^j(x)=\delta_j^i-2\frac{x_j x^i}{|x|^2},\;\;\;\;\;\;P_{ijab}=\frac{1}{2}(\delta_{ia}\delta_{jb}+\delta_{ib}\delta_{ja})-
\frac{1}{d+1}\delta_{ij}\delta_{ab}.
\ee

Since the quadratic on-shell action is a divergent quantity one needs to consider both boundary and counterterms in order to properly compute the two point function. 
In the present case 
for $z=1$ the terms of the renormalized action which could contribute to quadratic order 
perturbatively in  the metric are\footnote{Note that we are using Euclidean signature for metric.
(see for example \cite{{Shaghoulian:2015dwa},{Dehghani:2015gza}})}
\be
S_{\rm total}=S-\frac{1}{8\pi G}
\int d^{d+1}x \sqrt{\gamma}  K-\frac{1}{8\pi G}\int d^{d+1}x \sqrt{\gamma}\;
\left(\frac{r_F}{r}\right)^{\frac{\theta}{d}}\; \frac{d_\theta}{L},
\ee
where $S$ is the original action \eqref{action}. To evaluate the quadratic action it is also useful to note 
\be
\int d^{d+1}x\sqrt{\gamma}\; K=\partial_n\int  d^{d+1}x\sqrt{\gamma}
=\frac{r}{L}
\left(\frac{r_F}{r}\right)^{\frac{\theta}{d}}\partial_r\int d^{d+1}x\sqrt{\gamma},
\ee
with
\be
\sqrt{\gamma}=
\left( \frac{L}{r}
\right)^{d+1}\left(\frac{r}{r_F}\right)^{\theta+\frac{\theta}{d}}
\left(1+\frac{1}{2}h-\frac{1}{4}h^i_j h^j_i+\frac{1}{8}h^2+\cdots\right).
\ee
By plugging the linearized solution back into the action one finds (see
 \cite{Liu:1998bu} for more details)
\be
S_{\rm total}=\frac{1}{4}\frac{L^d}{16\pi G}\;\frac{d+2}
{d}\;\frac{\Gamma(d+2)}{\pi^{\frac{d+1}{2}}\Gamma\left( 
\frac{1+d}{2}\right)}\int d^{d+1}x\;d^{d+1}y\; \frac{h_{ab}(x)G_{abcd}(x,y)h_{cd}(y)}
{(x-y)^{2(d+1)}},
\ee
where $G_{abcd}(x,y)=J_a^i(x-y)J^b_j(x-y) P_{ijcd}$. 
Having found the quadratic on-shell action the  two point function of the  energy momentum tensor 
can be found as follows
\be
\langle T_{ab}(x) T_{cd}(y)\rangle=\frac{C_T}{(x-y)^{2(d+1)}}G_{abcd}(x,y).
\ee
where 
\be
C_T=\frac{L^d}{8\pi G}\;\;\frac{d+2}
{d}\;\frac{\Gamma(d+2)}{\pi^{\frac{d+1}{2}}\Gamma\left( 
\frac{1+d}{2}\right)},
\ee
For $z=1$ and $\theta\neq 0$ one can still find a solution for the equation of motion and 
evaluate the quadratic action. In this case going through the all steps mentioned above, one
arrives at 
\be
C_T=\frac{L^d}{8\pi Gr_F^{d-d_\theta}}\;\frac{d+2}{d}\;\frac{\Gamma(d_\theta+2)}
{\pi^{\frac{d+1}{2}}\Gamma\left( 
\frac{1+2d_\theta-d}{2}\right)}.
\ee 
It is worth noting that the  above expression may also be found 
from {the fact that} the equations of motion of metric perturbations in traceless-transverse gauge 
reduce to the equation of motion for a scalar field and therefore the 
corresponding two point function may be read from {the one of a} scalar field \cite{Dong:2012se}.


For $z\neq 1$, although it is not possible to find holographically the general form of the  
two point function of $T_{\mu\nu}$, we may still have a chance to compute the 
\textit{equal time} correlator. Although  we have not gone through the details of this idea, but from the analogous results of the scalar field \cite{Dong:2012se} one might expect to get the following expression 
\be
C_T\propto \frac{L^d}{8\pi Gr_F^{d-d_\theta}}\frac{\Gamma(d_\theta+z+1)}
{\pi^{\frac{d+1}{2}}\Gamma\left( 
\frac{2z-1+2d_\theta-d}{2}\right)}.
\ee
{We see} that here, differently from the holographic entanglement entropy, the coefficient does in 
fact depend on the Lifshiz exponent $z$.

\section{Explicit expressions for $\varphi_{2i}$ and $a_{2i}$ for $i=1,2,3$}\label{App:AppendixB}

In this appendix we will present the explicit form of the coefficients $\varphi_{2i}$ for 
the first few orders.
To proceed  let us start with the following series {Ansatz} for $\varphi$  
\be
\varphi(h)=\Omega+\varphi_2 h^2+\varphi_4 h^4+\varphi_6 h^6+{\cal O}(h^8).
\ee
Plugging this series in the equation of motion of $\varphi$ one arrives at the equation \eqref{eqphi} which can be solved order by {order}.  Doing so one finds
\bea
&&\varphi_2=-\frac{n \cot\Omega }{2 (d_\theta-1)},\;\;\;\;
\varphi_4=-\frac{n \cot\Omega [ (-2n+(d_\theta-1)^2)n \cot^2\Omega+(d_\theta-1)^2 
(6-2 d_\theta+n)]}{8 (d_\theta-3) (d_\theta-1)^3},\cr &&\cr &&
\varphi_6=-
\frac{8 (d_\theta+2) n^2-22 (d_\theta-1)^2 n+(3 d_\theta-7) (d_\theta-1)^3 }{48 (d_\theta-5) (d_\theta-3) (d_\theta-1)^5}\;n^3 \cot^5\Omega
\cr &&\cr
&&
\;\;\;\;\;\;\;\;\;
-\frac{2 (d_\theta (d_\theta+3)-20) n-(3 d_\theta-13) (d_\theta-1)^2-11 n^2}{24 (d_\theta-5) (d_\theta-3) (d_\theta-1)^3}\;n^2 \cot^3\Omega
\cr &&\cr &&
\;\;\;\;\;\;\;\;\;-\frac{(2 d_\theta-n-6) (4 d_\theta-n-20) }{48 (d_\theta-5) (d_\theta-3) (d_\theta-1)}\;n
\cot\Omega. 
\eea
It is clear from these expressions that the solution breaks down
for $d_\theta=2k+1$, $k=0,1,\cdots$. In this case one needs to modify the 
Anstatz by adding a {logarithmic} term.  For example for $d_\theta=3$, using the Ansatz
\be
\varphi(h)=\Omega+\varphi_2 h^2+\varphi_4 h^4  \left(c+\frac{1}{2}\log h^2\right)+
{\cal O}( h^6),
\ee
one finds\footnote{See also \cite{Myers:2012vs}}
\be
\varphi_2=-\frac{n}{4} \cot\Omega,\;\;\;\;\;\;\varphi_4=-\frac{n^2}{64} \left(n-4+n \cos2 \Omega \right)
\cot\Omega \csc^2\Omega,
\ee
where  $c$ remains unfixed. Similarly for {$d_\theta=5$} for the Ansatz
\be
\varphi(h)=\Omega+\varphi_2 h^2+\varphi_4 h^4+\varphi_6 h^6 \left(c+\frac{1}{2}\log h^2\right)+{\cal O}(h^8)
\ee
one arrives at
\bea
&&\varphi_2=-\frac{n}{8} \cot\Omega,\;\;\;\;\;\;\varphi_4=\frac{n}{512}   [(n-8) n 
\cot ^2\Omega-8 (n-4)]\cot\Omega,\cr &&\cr
&&\varphi_6=\frac{(n-4) (7 n-16)\; n\cot^4\Omega-4 (n (11 n-40)+32) 
\cot^2\Omega +32 (n-4)}{12288}\;n^2 \cot\Omega, 
\eea
with unspecified $c$.

Having found the coefficients $\varphi_{2i}$ it is straightforward to find 
the coefficients $a_{2i}$ appearing in the equation \eqref{eq area integrand expansion hs}.
The results are 
\bea
a_0 \!\!&\!\!=\!\!&\!\!\sin^{n}\!\Omega,\;\;\;\;\;\;\;\;\;\; a_2 =  \varphi_2 (2 \varphi_2+n \cot\Omega )
\sin^n\Omega
\\
a_4 \!\!&\!\!=\!\!&\!\! \frac{1}{2}[n \left(2 \varphi_2^3+\varphi_4\right) 
\sin2 \Omega-\varphi_2 \sin^2\Omega \left(\varphi_2 \left(4 \varphi_2^2+n-4\right)-16 \varphi_4\right)
+\varphi_2^2 (n-1) n \cos^2\Omega ]\sin^{n-2}\!\Omega .\nonumber
\eea
Note that for the particular values of $d_\theta=1,3$ one needs to use the proper results
of $\varphi_{2i}$ given in this appendix.

\end{document}